\documentclass[aip,jcp,reprint,amsmath,amssymb,floatfix,longbibliography]{revtex4-2}
\usepackage[utf8]{inputenc}
\usepackage[T1]{fontenc}
\usepackage[USenglish]{babel}

\usepackage{upgreek}
\usepackage[normalem]{ulem}

\usepackage{xcolor}
\usepackage{graphicx}
\usepackage{tabularx}
\usepackage{amsfonts}
\usepackage{mathtools}

\usepackage{xcolor}
\usepackage{comment}
\usepackage{hyperref}
\usepackage{cleveref}

\usepackage{booktabs,cancel,colortbl,csquotes,helvet,mathpazo,listings,pgfplots}
\pgfplotsset{compat=1.18}

\usepackage[range-phrase={\text{~to~}},output-decimal-marker={.}]{siunitx}
\DeclareSIUnit\photons{\textrm{ph}}

\usepackage[colorinlistoftodos,textsize=small]{todonotes}
\setuptodonotes{fancyline, color=green!40}
\usepackage[version=4]{mhchem}
\usepackage{lipsum}

\makeatletter
\renewcommand{\fnum@figure}{\textbf{\figurename~\thefigure}}
\renewcommand{\fnum@table}{\textbf{\tablename~\thetable}}
\makeatother

\usepackage{multirow}   

\begin{document}

\date{\today}

\title{Liquid-liquid phase separation precedes crystallization in supercooled water-glycerol solutions}

\author{Iason Andronis}
\thanks{These authors contributed equally to this work.}
\affiliation{Department of Physics, AlbaNova University Center, Stockholm University, 106 91 Stockholm, Sweden}
\author{Sharon Berkowicz}
\thanks{These authors contributed equally to this work.}
\affiliation{Department of Physics, AlbaNova University Center, Stockholm University, 106 91 Stockholm, Sweden}
\author{Mariia Filianina}
\thanks{Current affiliation: Department of Physics, Saint Petersburg State University, Saint Petersburg 198504, Russia}
\affiliation{Department of Physics, AlbaNova University Center, Stockholm University, 106 91 Stockholm, Sweden}
\author{Maddalena Bin}
\affiliation{Department of Physics, AlbaNova University Center, Stockholm University, 106 91 Stockholm, Sweden}
\author{Robin Tyburski}
\affiliation{Department of Physics, AlbaNova University Center, Stockholm University, 106 91 Stockholm, Sweden}
\author{Robert Bauer}
\affiliation{Deutsches Elektronen-Synchrotron DESY, Notkestr. 85, 22607 Hamburg, Germany}
\affiliation{Freiberg Water Research Center, Technische Universität Bergakademie Freiberg, 09599 Freiberg, Germany}
\author{Yuriy Chushkin}
\affiliation{ESRF -- The European Synchrotron, F-38043, Grenoble, France}
\author{Federico Zontone}
\affiliation{ESRF -- The European Synchrotron, F-38043, Grenoble, France}
\author{Michael Sprung}
\affiliation{Deutsches Elektronen-Synchrotron DESY, Notkestr. 85, 22607 Hamburg, Germany}
\author{Wojciech Roseker}
\affiliation{Deutsches Elektronen-Synchrotron DESY, Notkestr. 85, 22607 Hamburg, Germany}
\author{Fabian Westermeier}
\affiliation{Deutsches Elektronen-Synchrotron DESY, Notkestr. 85, 22607 Hamburg, Germany}
\author{Felix Lehmkühler}
\affiliation{Deutsches Elektronen-Synchrotron DESY, Notkestr. 85, 22607 Hamburg, Germany}
\author{Fivos Perakis}
\email{f.perakis@fysik.su.se}
\affiliation{Department of Physics, AlbaNova University Center, Stockholm University, 106 91 Stockholm, Sweden}

\begin{abstract}
Understanding the structural evolution of supercooled water-glycerol solutions is important for cryopreservation, yet distinguishing liquid-state transformations from ice crystallization remains challenging. Here, we investigate a deeply supercooled water-glycerol solution by X-ray photon correlation spectroscopy (XPCS) in ultra-small-angle X-ray scattering (USAXS) geometry, combined with wide-angle X-ray scattering (WAXS). This combination simultaneously captures the structural and dynamical evolution of the supercooled liquid upon quenching to cryogenic temperatures (172~K). We observe discontinuous changes in the liquid structure on molecular length scales and formation of microscale domains. The dynamics slow down during this stage and exhibit hyper-diffusive, ballistic-like relaxation.
This transformation precedes ice crystallization, which we identify from the emergence of ice Bragg peaks in WAXS, allowing the two processes to be temporally separated. Phase-field (Cahn-Hilliard) simulations qualitatively reproduce the experimental observations and show that a spinodal-decomposition scenario is consistent with the measured scattering evolution. These findings are consistent with a liquid-liquid phase separation scenario preceding ice crystallization and provide a route to disentangle the two processes in supercooled aqueous systems.
\end{abstract}

\maketitle

\section{Introduction}\label{s:introduction}
Although liquids are traditionally viewed as having a single thermodynamic state, growing evidence suggests that some materials can exist in multiple distinct liquid or amorphous forms, a phenomenon known as polyamorphism~\cite{stanley_liquid_2013}. Such behavior is particularly relevant in metastable liquids, including deeply supercooled water, where theory and experiment have suggested the existence of a liquid-liquid transition (LLT) between high- and low-density liquid states. This transition has been proposed to originate from an underlying liquid-liquid critical point based on evidence both from simulations ~\cite{debenedetti_second_2020} and experiments~\cite{perakis_diffusive_2017, kim_experimental_2020,you_experimental_2026}. 

It is therefore of interest to explore whether the LLT of water also manifests in aqueous solutions, in which the solutes suppress ice crystallization and thereby provide access to deeper supercooling. Typical examples include water-glycerol~\cite{murata_liquid_2012, murata_general_2013} or trehalose~\cite{suzuki_direct_2022} solutions, which are widely used as cryoprotectants. Previous structural and kinetic studies of water-glycerol mixtures have reported signatures consistent with a low-temperature liquid transformation~\cite{murata_liquid_2012,murata_general_2013}. However, whether these observations originate from a genuine liquid-liquid transformation or from crystallization-induced processes, such as freeze-concentration or solute demixing, remains debated because both occur in the same temperature regime~\cite{murata_liquid_2012, suzuki_experimentally_2014, popov_puzzling_2015, bachler_glass_2016, jahn_glass_2016, alba-simionesco_interplay_2022}. Resolving this question requires measurements that can directly distinguish structural changes in the liquid from the onset of ice formation.

Studies of glassy water-glycerol mixtures suggest that high- and low-density amorphous (HDA and LDA) ice-like states cease to exist above approximately 5~mol\%~\cite{bachler_glass_2016, jahn_glass_2016} glycerol. More recently, femtosecond X-ray scattering and molecular dynamics simulations of dilute (3.2~mol\%  glycerol) water-glycerol microdroplets revealed that isothermal compressibility increases upon supercooling, and exhibits a maximum~\cite{berkowicz_supercritical_2024}. This maximum is associated with the existence of a Widom line~\cite{xu_relation_2005, kim_maxima_2017} emanating from a critical point. The addition of glycerol shifts that maximum from 230~K to 223~K, indicating a shift of the solution critical point to lower temperatures and higher pressures~\cite{berkowicz_supercritical_2024}.

\begin{figure*}[htb]
\centering
 \includegraphics[width=\textwidth]{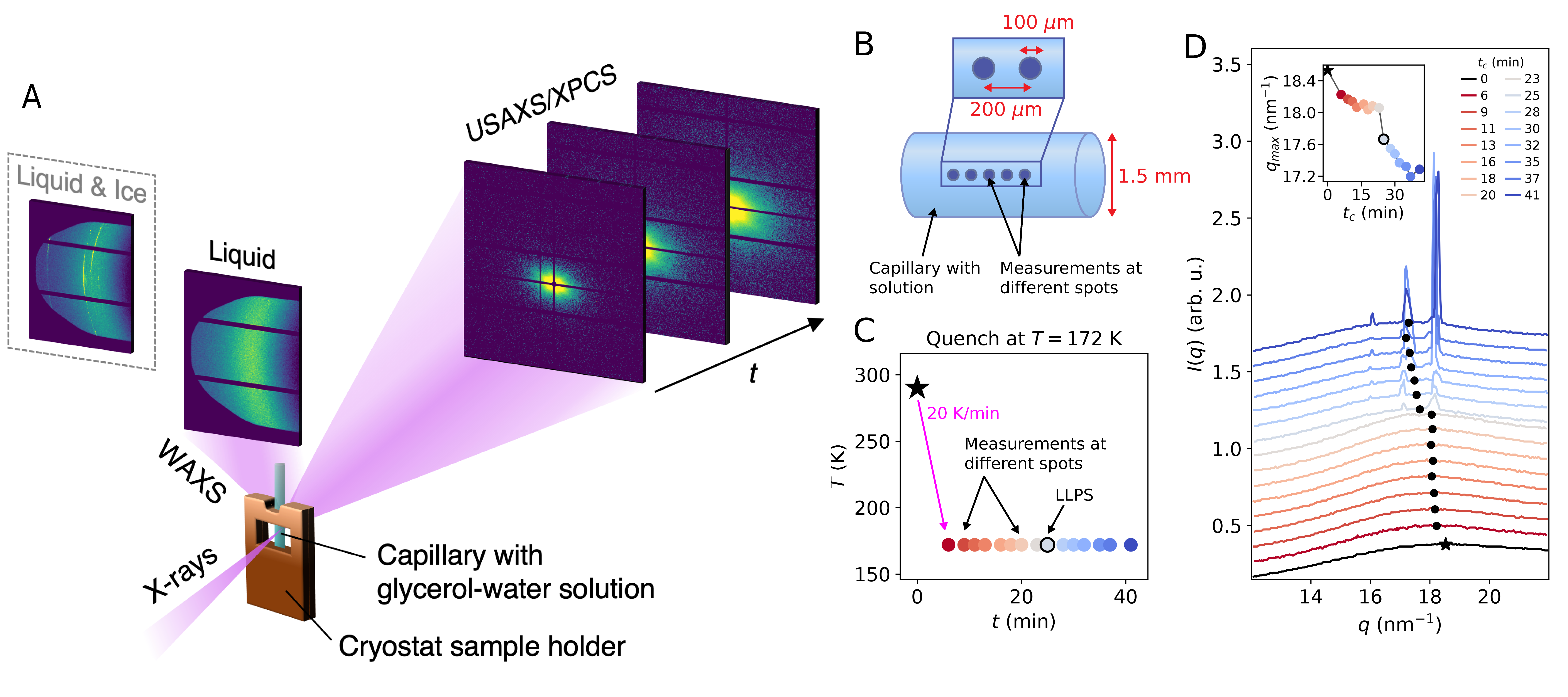}
 \caption{\label{fig:setup_protocol}
Experimental setup and measurement protocol for simultaneous WAXS, USAXS, and XPCS measurements of supercooled water-glycerol solution. (A) Schematic of the experimental setup. The 16.5 mol\% water-glycerol solution was quenched to $T=172\pm6$~K in a liquid nitrogen-cooled cryostat and measured simultaneously by wide-angle X-ray scattering (WAXS) and X-ray photon correlation spectroscopy (XPCS) in the ultra-small-angle X-ray scattering (USAXS) geometry. WAXS probes changes in the liquid intermolecular structure and the onset of ice crystallization, whereas USAXS/XPCS probes mesoscale structure and dynamics. (B) Schematic of a capillary demonstrating the measurements in different sample positions. (C) Temperature protocol. One measurement was recorded before quenching (black star), followed by 130 s measurements at different cooling times (filled circles). The black circle marks the measurement after 25 min of cooling analyzed throughout the remainder of the paper. (D) WAXS patterns recorded during cooling. The inset shows the evolution of the liquid peak position $q_{\max}$ with cooling time.
 }
\end{figure*}

Here, we explore supercooled water-glycerol solutions by combining X-ray photon correlation spectroscopy (XPCS) in the ultra-small-angle scattering (USAXS) regime with wide-angle X-ray scattering (WAXS). This approach simultaneously probes both structural and dynamical information at nanometer length scales (from USAXS/XPCS), while monitoring the changes in the liquid intermolecular structure and the onset of ice formation (from WAXS). A schematic of the experiment is shown in Fig.~\ref{fig:setup_protocol}A. In addition, to help interpret the experimental observations, we also performed phase-field simulations using the Cahn-Hilliard (CH) framework~\cite{cahn_free_1958, cahn_phase_1965, sciortino_interference_1993}. The CH model describes qualitatively the experimental observations and shows that a spinodal-decomposition scenario is consistent with the measured structural evolution prior to crystallization.

\section{Methods}\label{s:methods}

\subsection{Experimental details}

Experiments were performed at beamline P10 (PETRA III, DESY) and ID10 (ESRF). This manuscript reports P10 results, while analogous measurements at ID10 are consistent with this analysis. The sample, 16.5~mol\% water-glycerol solution, was measured in a 1.5-mm thick sealed quartz capillary mounted inside a liquid-nitrogen cooled cryostat (see Fig.~\ref{fig:setup_protocol}A). We developed a measurement scheme in which the sample was quenched from room temperature ($T=295$~K), at a cooling rate of 20~K/min, to a constant target temperature. The sample temperature after quenching was calibrated based on the melting point of acetone ($T_m=178$~K) and water. The simultaneous WAXS/USAXS/XPCS measurements were initiated at room temperature prior to the quench, which is defined as the time zero of cooling time. Once the target temperature was reached, data were collected in a series of measurements in different sample spots (200~$\mu$m center-to-center separation, twice the beam size, to ensure we do not illuminate the same spot twice) over the course of ${\approx}40$~min (see Fig.~\ref{fig:setup_protocol}B and C). The measurement time in each sample spot was limited to 130~s (10,000~scattering patterns with 0.013~s exposure time) which we define as the waiting time $t$. \@WAXS data were measured using a Pilatus 300k detector positioned close to the sample, while USAXS/XPCS data were measured with an Eiger X4M detector at a sample-to-detector distance of 21.2~m. To obtain azimuthally integrated intensity profiles $I(q)$ as a function of scattering vector $q = \frac{4\pi}{\lambda}\sin\theta$, we performed 2D detector integration using the Python library \textsc{pyFAI}~\cite{ashiotis_fast_2015}. The resulting $q$ range in WAXS is $q=12-22$~nm$^{-1}$ while for USAXS $q=0.008-0.2$~nm$^{-1}$. We used a monochromatic X-ray beam with photon energy 9.0~keV, including the use of an Si(111) monochromator, and a square beam with a size (width) of $100~\mu\mathrm{m}$. With the insertion of Si attenuators, the flux on the sample was $\approx 1.5 \times 10^{10}$~ph/s. These experimental details are outlined in Table~\ref{tab:exp-param}. 

\begin{table}[htb]
  \centering
  \caption{Experimental parameters for the X-ray setup and sample environment.
  }\label{tab:exp-param}
  \begin{center}
    \begin{tabular}{lcc}
        \toprule
        \textbf{Measurement}                 & \textbf{WAXS}             & \textbf{USAXS/XPCS} \\
        \midrule
        \textbf{Detector}                    & Pilatus 300k              & Eiger X4M \\
        \textbf{q-range}                     & 12--22~nm$^{-1}$          & 0.008--0.2~nm$^{-1}$ \\
        \textbf{Sample-to-detector distance} & 0.31~m                    & 21.2~m \\
        \textbf{Exposure time}               & \multicolumn{2}{c}{0.013~s} \\
        \textbf{Photon energy}               & \multicolumn{2}{c}{9.0~keV} \\
        \textbf{Square beam size}               & \multicolumn{2}{c}{$100$~$\mu$m} \\
        \textbf{Flux on sample}              & \multicolumn{2}{c}{$1.5\times10^{10}$~ph/s} \\
        \textbf{Sample environment}          & \multicolumn{2}{c}{LN$_2$ cooled cryostat} \\
      \bottomrule
    \end{tabular}
  \end{center}
\end{table}


The measurement protocol is illustrated in Fig.~\ref{fig:setup_protocol}B and C. Following the quench (with 20~K/min), a series of 130~s measurements was recorded in different sample spots as a function of cooling time. Figure~\ref{fig:setup_protocol}D shows the evolution of the WAXS patterns as a function of cooling time upon quenching to $T_{quench}=172\pm6$~K. Upon cooling, the liquid peak position shifts rapidly to lower $q$ until 6~min of cooling, then more slowly as the sample temperature equilibrates and settles at $q_{\max}=18.1$~nm$^{-1}$ within 15--20~min (Fig.~\ref{fig:setup_protocol}D, inset). During this cooling time, there is evidently no sign of Bragg peaks from crystallization in the WAXS patterns, indicating that the supercooled solution is still in the liquid state. The first Bragg peaks, indicating the growth of ice crystallites, emerge towards the end of the measurement at 25~min of cooling. In the remainder of the paper, we analyze the kinetics and dynamics during the evolution between the data series obtained just before the onset of Bragg peaks.

\begin{figure*}[htb]
\centering
 \includegraphics[width=1\textwidth]{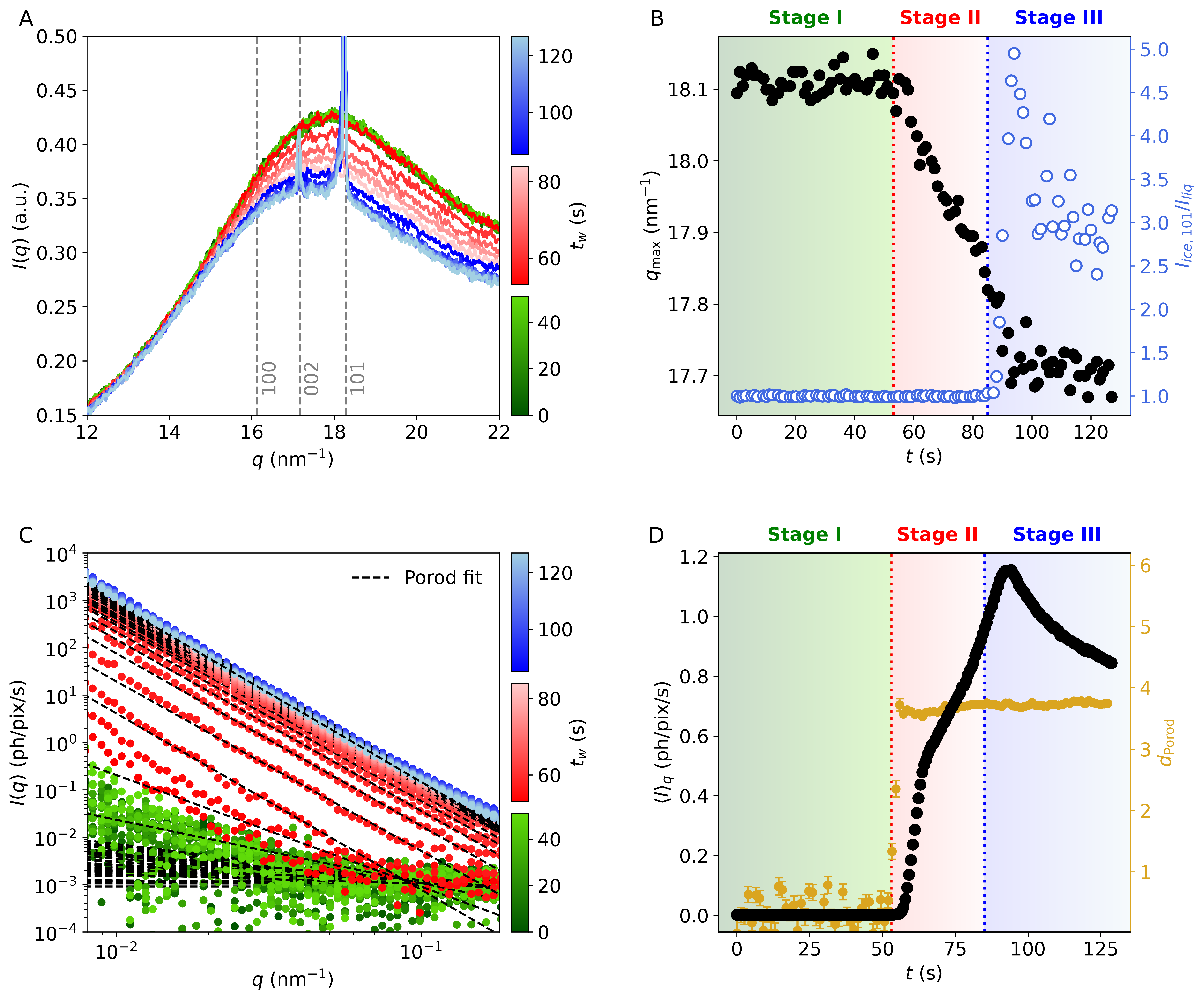}
 \caption{\label{fig:WAXS_saxs} Simultaneous WAXS (A, B) and USAXS (C, D) measurements during the structural transformation at $T=172\pm6$~K after 25~min of cooling. Three distinct stages are identified: a stable supercooled liquid (stage I, green), a rapid structural transformation (stage II, red), and the onset of ice crystallization (stage III, blue). (A) WAXS patterns $I(q)$ versus waiting time $t$ (colorbar); vertical dashed lines mark the (100), (002) and (101) Bragg peaks of hexagonal ice. (B) Evolution of the liquid peak position $q_{\max}$ (black circles, left axis) and the intensity ratio $I_{ice,101}/I_{liq}$ of the ice (101) Bragg peak at $q=18.26$~nm$^{-1}$ to the liquid peak (open blue circles, right axis) versus $t$. The dotted lines and shaded regions indicate the boundaries between the different stages. (C) USAXS patterns $I(q)$ versus $t$. Black dashed lines show power-law (Porod) fits to Eq.~\eqref{eq:saxs_fit}. (D) Mean scattering intensity $\langle I\rangle_q$ averaged over $0.01\leq q\leq0.2$~nm$^{-1}$ (black, left axis) and the fitted Porod exponent $d_{Porod}$ from the fits in panel~C (gold, right axis) versus $t$.
 }
\end{figure*}

\subsection{Phase-field simulation details}\label{ss:ch}
We modeled the phase separation within the Cahn-Hilliard (CH) framework~\cite{cahn_free_1958, cahn_phase_1965, sciortino_interference_1993} as a minimal model that contrasts spinodal decomposition (SD) and nucleation-and-growth (NG). The time evolution of a conserved order parameter $\phi(\mathbf{r},t)\in[-1,1]$ follows the CH equation $\partial_t\phi=\nabla\cdot(M\nabla\mu)$, where $M$ is the mobility and the chemical potential $\mu=\delta F/\delta\phi$ is the variational derivative of a coarse-grained free-energy functional $F[\phi] = \int[f(\phi) + \kappa|\nabla\phi|^2]\,dV$. We employ the standard mean-field (Ginzburg---Landau) form of $F[\phi]$: a local double-well bulk free-energy density $f(\phi)=\tfrac{1}{4}\phi^4-\tfrac{1}{2}\epsilon\phi^2$, with $\epsilon=(T_c-T)/T_c$ being the reduced temperature, supplemented by the square-gradient penalty term $\kappa|\nabla\phi|^2$ that accounts for the interfacial energetic cost of spatial variations in the order parameter. We interpret $\phi(\mathbf{r},t)$ as a coarse-grained, normalized density field of the water-glycerol solution, such that spatial variations in $\phi$ represent mesoscale density fluctuations that develop during the phase separation. The mapping between $\phi$ and an absolute mass density is not specified, since the simulations serve as a minimal model for the pathway and relative evolution of density heterogeneities rather than for quantitative, length-scale-calibrated predictions.

We solved the CH equation using the semi-implicit Fourier-spectral scheme~\cite{chen_applications_1998}, which mitigates the severe stability constraints of explicit schemes for the fourth-order CH dynamics while remaining efficient and spectrally accurate. The scattering intensity $I(q,t)$ was obtained from the radially averaged spectral intensity of the order-parameter field, computed via its Fourier transform. The representative SD run (Fig.~\ref{fig:CH_summary}A) was initialized with $\phi=0.2$ and a small initial noise amplitude ($0.01$), the NG run (Fig.~\ref{fig:CH_summary}B) with $\phi=0.6$ and a larger initial noise amplitude ($2$) to promote nucleation-type behavior, both at reduced temperature $T=0.2\,T_c$.

\section{Results}\label{s:results}

The water-glycerol solution exhibits a WAXS peak, which relates to the liquid intermolecular structure, centered around $q_{\max}\approx18.5$~nm$^{-1}$ at room temperature $T=295$~K (see Fig.~\ref{fig:setup_protocol}D). Upon quenching to $T=172\pm6$~K, the WAXS peak shifts to lower $q$ and stabilizes at $q_{\max}\approx18.1$~nm$^{-1}$ within 15--20~min of cooling, indicating that the sample has reached temperature equilibrium. Figure~\ref{fig:WAXS_saxs}A displays the temporal evolution of the WAXS pattern at $T=172\pm6$~K after 25~min of cooling. The WAXS patterns as a function of the waiting time $t$ are colored according to three observed stages of liquid transformation. These stages are characterized by an initially constant WAXS peak position at $q_{\max}\approx18.1$~nm$^{-1}$ (stage I, in green), a sudden and rapid shift of the WAXS peak position towards lower $q$ starting at $t\approx52$~s (stage II, in red), and the subsequent onset of ice crystallization observable by the emergence of ice Bragg peaks at $t\geq85$~s (stage III, in blue).

Figure~\ref{fig:WAXS_saxs}B (black circles, left axis) highlights the step-like shift of the WAXS peak position $q_{\max}$ to lower $q$, from $q_{\max}\approx18.1$~nm$^{-1}$ to $q_{\max}\approx17.7$~nm$^{-1}$. This shift initiates abruptly in stage II (red dotted line), but continues and finalizes alongside the crystallization process in stage III (from $t=85$~s, blue dotted line). 

Figure~\ref{fig:WAXS_saxs}B (open blue circles, right axis) also shows the temporal evolution of the relative intensity of the first observed ice Bragg peak at $q=18.26$~nm$^{-1}$ and the liquid peak in WAXS.\@ Here, the sharp rise in the relative intensity as ice crystallites emerge after $t=85$~s is clearly evident (blue dotted line). This increase of the Bragg peak intensity is also separate in time from the start of the WAXS peak shift (red dotted line), suggesting that these are two distinct processes, each with a sharp, discontinuous onset.

Using the Scherrer equation $\delta=K\lambda/(\Delta\cos{(\theta)})$, where $\lambda$ is the X-ray wavelength, $\theta$ is the scattering angle and $\Delta$ is the Bragg peak full-width at half-maximum (FWHM), we estimate the initial average crystallite size $\delta$ to be $\delta\approx30-40$~nm. We cannot fully exclude the presence of a small population of nanocrystallites at earlier times ($t<85$~s). X-ray scattering can detect very small crystalline fractions; for example, ice fractions as low as $\approx0.05\%$ by mass have been detected in evaporatively cooled water droplets, assuming a crystallite size of 12 nm~\cite{sellberg_ultrafast_2014}. Nevertheless, a small number of nanocrystallites could remain undetected if they produced isolated Bragg spots at azimuthal angles outside the WAXS detector coverage. Importantly, however, the ice Bragg intensity remains at the baseline throughout stage II and then increases sharply at $t\approx85$~s. This abrupt onset is difficult to reconcile with a gradual accumulation of crystallites during stage II and instead indicates that detectable crystallization begins only after the preceding structural and dynamical transformation. 
Even if isolated crystallites were present below the detection limit, they would have to induce abrupt changes in the liquid structure, mesoscale scattering, and nanoscale dynamics essentially simultaneously. Such coordinated behavior is not expected from the gradual accumulation of a sparse population of nanocrystallites.

The ice Bragg peaks in stage III intensify over time (see Fig.~\ref{fig:setup_protocol}D), suggesting that the ice crystallites continue to nucleate and grow. Eventually, the WAXS patterns exhibit all three characteristic Bragg peaks of hexagonal ice ($I_h$ 100, 002 and 101) within the accessible WAXS $q$-range. Along with the continued freezing at longer cooling time, the liquid peak position in WAXS continues to shift towards lower $q$ more gradually, at about an order of magnitude slower rate than the initial rapid shift in stage II (see Fig.~\ref{fig:setup_protocol}D, inset). This slower shift is consistent with a continuously increasing glycerol fraction in the liquid phase due to freeze-concentration, during which glycerol molecules are expelled from growing ice crystallites. 

Figure~\ref{fig:WAXS_saxs}C shows the temporal evolution of the corresponding USAXS patterns at 25~min of cooling. Here, we observe a strong enhancement of the USAXS intensity (stage II, in red), which occurs concurrently with the rapid changes in WAXS (Fig.~\ref{fig:WAXS_saxs}A,B). The sudden rise of the $q$-averaged USAXS intensity $\langle I\rangle_q$ is displayed in Fig.~\ref{fig:WAXS_saxs}D (black, left axis). During stage III (marked in blue), while ice crystallites are present according to WAXS, the USAXS intensity exhibits a maximum around $t\approx100$~s. 
Notably, the maximum of $\langle I\rangle_q$ in USAXS is coincident with the crossover from the rapid shift of $q_{\max}$ to the lower plateau value at $q_{\max}=17.7$~nm$^{-1}$ in WAXS (see Fig.~\ref{fig:WAXS_saxs}B). The simultaneous occurrence of these two events is reproduced following the quenching of a fresh sample with identical glycerol concentration (16.5~mol\%) to $T=182\pm6$~K (see the supplementary material, Sec.~S2).

The USAXS patterns in Fig.~\ref{fig:WAXS_saxs}C are well described by a power-law function (black dashed lines in Fig.~\ref{fig:WAXS_saxs}C):

\begin{equation}
    I(q) = A q^{-d_{Porod}} ~,
    \label{eq:saxs_fit}
\end{equation}

where $A$ is a constant and $d_{Porod}$ is the power-law (Porod) exponent, related to the shape of the scattering domains. Generally, power-law-like scattering appears in the $q$-range that relates to length scales $\sim2\pi/q$ significantly smaller than the domain size. Figure~\ref{fig:WAXS_saxs}D (gold, right axis) shows the fit result of the USAXS patterns to Eq.~\ref{eq:saxs_fit}. It reveals a discontinuous, step-like increase of the power-law exponent from $d_{Porod}\approx1$ to $d_{Porod}\approx4$ upon entering stage II (red dotted line). A Porod exponent of $d_{Porod}=4$ in this $q$-range (corresponding to length scales $2\pi/q\approx0.1-1~\mu$m) is indicative of scattering from micrometer-sized domains with sharp interfaces~\cite{als-nielsen_elements_2011}. 

\begin{figure*}[htb]
\centering
 \includegraphics[width=1\textwidth]{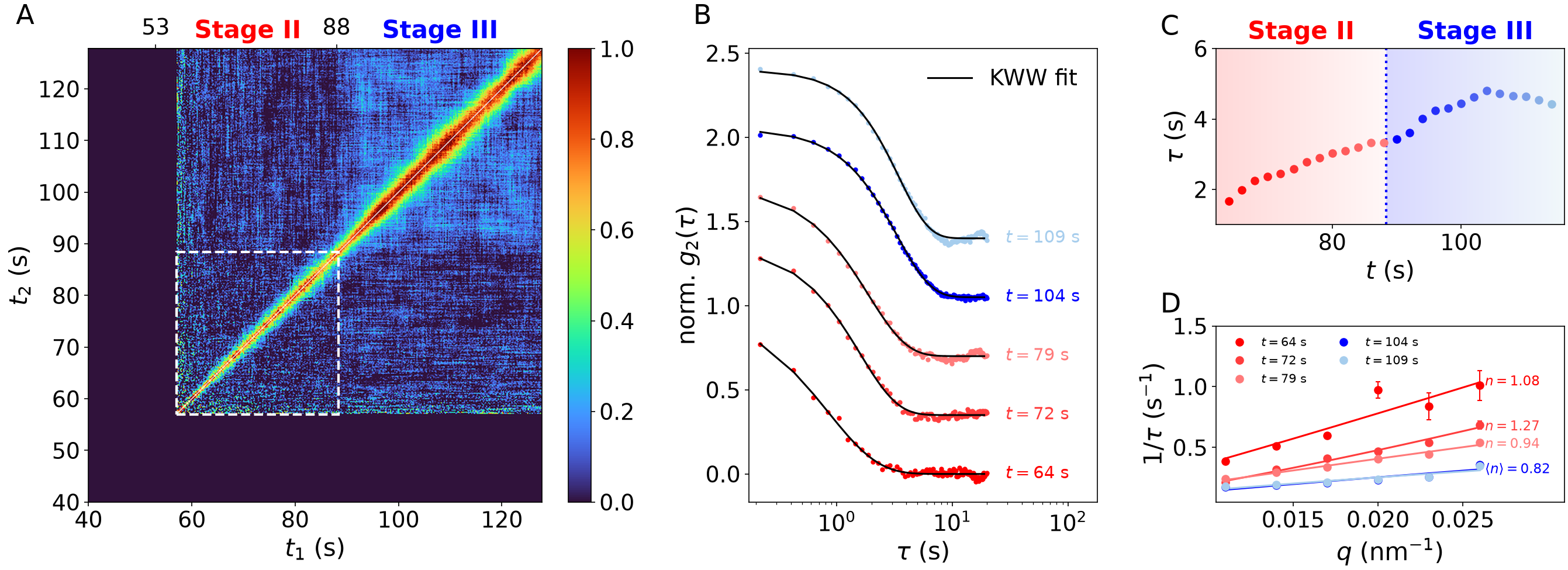}
 \caption{\label{fig:xpcs} XPCS measurements of the nanoscale dynamics during the structural transformation. (A) The two-time correlation function (TTC) at $q \approx 0.016$~nm$^{-1}$.
 The white dashed square indicates the part of the TTC that is compared with the simulations in Fig.~\ref{fig:CH_ttc}.
 (B) The intensity autocorrelation functions $g_2$ for different waiting times $t$ at $q=0.016$~nm$^{-1}$, extracted from the TTC.
 The experimental data (data points) are fitted to Kohlrausch-Williams-Watts (KWW) functions (black lines) and normalized by the fitted offset $c$ and contrast $\beta$. An offset has been introduced for each curve, to facilitate the comparison. (C) The average relaxation time $\tau$ obtained from the KWW fits, averaged over $0.013\leq q\leq 0.019$~nm$^{-1}$. (D) The $q$-dependence of the inverse relaxation time $1/\tau$. Solid lines depict $1/\tau \propto q^n$ fits, with the fitted exponent $n$ indicated in the figure. 
 Data acquired before and after the onset of ice crystallization ($t=85$~s) are shown in red and blue, respectively.
 }
\end{figure*}

Analysis of the two-time correlation function (TTC) from XPCS allows us to follow the dynamics during the aforementioned structural transitions in the supercooled solution. Specifically, the TTC~\cite{brown_speckle_1997,sutton_using_2003,madsen_simple_2010} (denoted $c_2$) is calculated as
\begin{equation}
    c_2(q,t_1,t_2) = \frac{\langle I(q,t_1) I(q,t_2) \rangle_{\text{pix}}}{\langle I(q,t_1)\rangle_{\text{pix}} \langle I(q,t_2)\rangle_{\text{pix}}} ~,
    \label{c2}
\end{equation}
where the brackets indicate averaging over detector pixels within an annulus of constant $q$. The $I(q,t_1)$ and $I(q,t_2)$ are the scattering intensities at times $t_1$ and $t_2$, respectively. The pixel average is restricted to the azimuthal sectors free of static anisotropic scattering, by means of a per-sample mask; this mask and the resulting speckle contrast are documented in the supplementary material, Sec.~S3. 
Figure~\ref{fig:xpcs}A shows a calculated TTC function (at $q=0.016$~nm$^{-1}$) for the time interval starting at 25~min of cooling. The TTC color scale is clipped to (1, 1.1) and rescaled to the range (0, 1), to enhance the visual contrast of the decorrelation dynamics against the flat, uncorrelated baseline. The temporal evolution of the dynamics can be followed as a function of the waiting time $t$ along the main diagonal of the TTC (i.e.\ along $t_1=t_2$ in Fig.~\ref{fig:xpcs}A). Here, we observe the appearance of a baseline in the TTC contrast at around $t\approx58$~s, which coincides with the rapid liquid peak shift in WAXS (Fig.~\ref{fig:WAXS_saxs}B) and the USAXS enhancement (Fig.~\ref{fig:WAXS_saxs}D) at times $t\geq 52$~s. 

The intensity autocorrelation function $g_2$ for different waiting times, i.e. the instantaneous dynamics, can be extracted from the $c_2$ correlation decays emanating from the main diagonal, i.e.
\begin{equation}
    g_2(q, t_1) = \langle c_2(q, t_1\geq t_2, t_2)\rangle_{t_2\in (t-\delta t, t+\delta t)} ~, 
    \label{g2}
\end{equation}
where the bracket notation indicates averaging over a narrow range of waiting times, $t_2\in (t-\delta t, t+\delta t)$, with $\delta t=5.12~$s. The value of $\delta t$ was selected so that the dynamics within the averaging range are static with respect to the measuring time $t$.
The $g_2$ functions at $q=0.016$~nm$^{-1}$ for different waiting times are presented in Fig.~\ref{fig:xpcs}B and are well described by Kohlrausch-Williams-Watts (KWW) exponential functions (solid lines):
\begin{equation}
   g_2(q,t_1) = \beta(q)\,\exp\!\left[ -2 {\left(t_1/\tau_0(q)\right)}^{\alpha(q)} \right] + c ~,
    \label{g2fit}
\end{equation}
where $\beta$ is the speckle contrast, $\tau_0$ is the relaxation time, $\alpha$ is the KWW exponent and $c$ is the offset. The average relaxation time is given by $\tau=\frac{\tau_0}{\alpha}\Gamma(1/\alpha)$ and $\Gamma$ is the Gamma function~\cite{guo_entanglementcontrolled_2012}. The $g_2$ fits were restricted to the time window preceding the onset of baseline distortions associated with the evolving TTC. Accordingly, only the fitted portion of each $g_2$ curve is displayed.

Figure~\ref{fig:xpcs}C shows the temporal evolution of the relaxation time $\tau$ from fits to Eq.~\eqref{g2fit}, averaged over $0.013\leq q\leq0.019$~nm$^{-1}$. Evidently, there is a gradual slowing down in the dynamics as the transition proceeds in stage II (red, $t\leq85$~s). The relaxation time increases from $\tau \approx 2$~s to $\tau\approx 4-5~$s, while it gradually plateaus along with the formation of ice crystallites in stage III (blue, $t\geq85$~s).

In addition, we find that the KWW exponent ranges between $1<\alpha<2$; $\alpha\approx1.1-1.2$ during the initial stage of the transition (stage II) while $\alpha\approx1.5$ under the influence of ice crystallites (stage III). The time dependence of the $q$-averaged KWW exponent can be found in the supplementary material, Sec.~S1.1. We further analyze the $q$-dependence of the relaxation rate by fitting $1/\tau\propto q^n$ (Fig.~\ref{fig:xpcs}D, solid lines). Averaging the fitted exponents over the waiting times within each stage yields $n=1.08\pm0.13$ in stage II and $n=1.01\pm0.08$ in stage III. 
The fitting procedure and the ROI selection are described in the supplementary material, Sec.~S1.2. Together with the compressed-exponential line shape ($\alpha>1$), a relaxation rate $1/\tau\propto q$ corresponds to hyper-diffusive, ballistic-like motion, which we observe in both stages.
Such hyper-diffusive dynamics have been associated with SD-type LLPS dynamics in soft matter systems, including protein solutions~\cite{girelli_microscopic_2021} and colloidal gels~\cite{gao_microdynamics_2015}, where they were assigned to domain growth and coarsening during phase separation. Whether it is indeed LLPS that is giving rise to this dynamic behavior in the present case, as well as the associated structural changes presented above, will be discussed in Sec.~\ref{s:discussion}. 

\begin{figure*}[htb]
    \centering
    \includegraphics[width=1\textwidth]{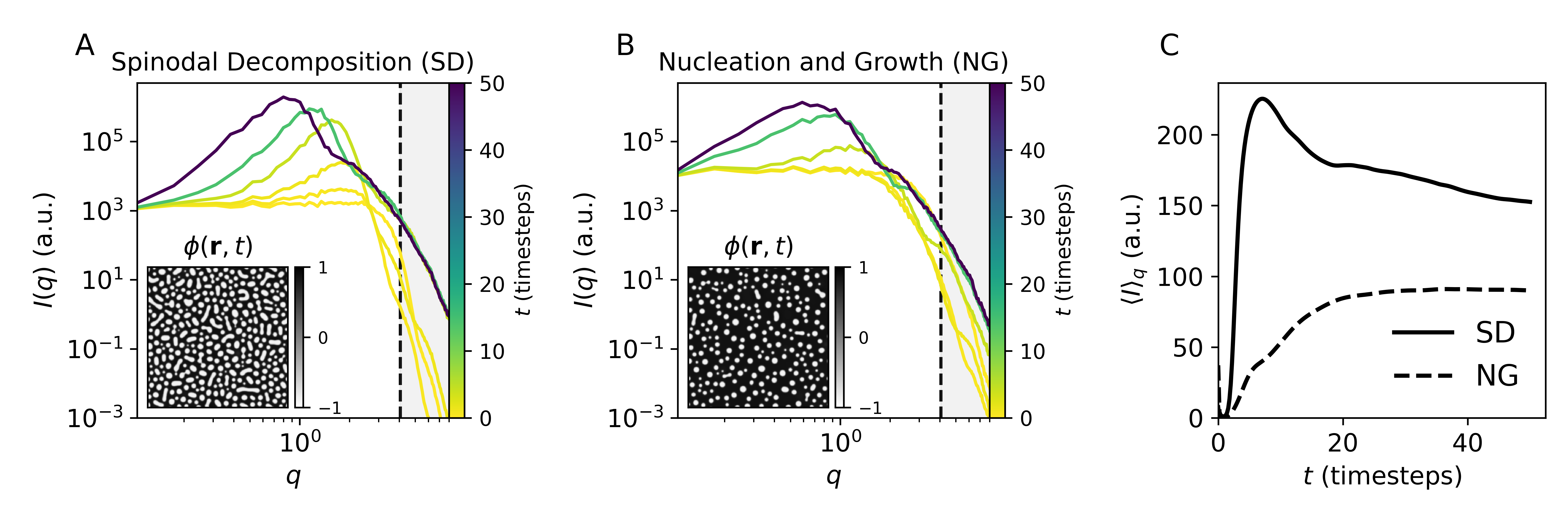}
    \caption{Structural evolution obtained from Cahn-Hilliard (CH) simulations. (A) Simulated scattering intensity $I(q)$ for a representative spinodal decomposition (SD) simulation. The momentum transfer is reported in reciprocal simulation units. The shaded region indicates the q-range over which $\langle I \rangle_q$ is calculated. (B) Simulated scattering intensity for a representative nucleation-and-growth (NG) simulation. The inset shows the corresponding order-parameter field $\phi(\mathbf{r},t)$ at $t=45$ timesteps. (C) Mean scattering intensity $\langle I\rangle_q$ extracted from the shaded regions in panels A and B as a function of simulation time $t$.}\label{fig:CH_summary}
\end{figure*}

\begin{figure*}[htb]
    \centering
    \includegraphics[width=0.95\textwidth]{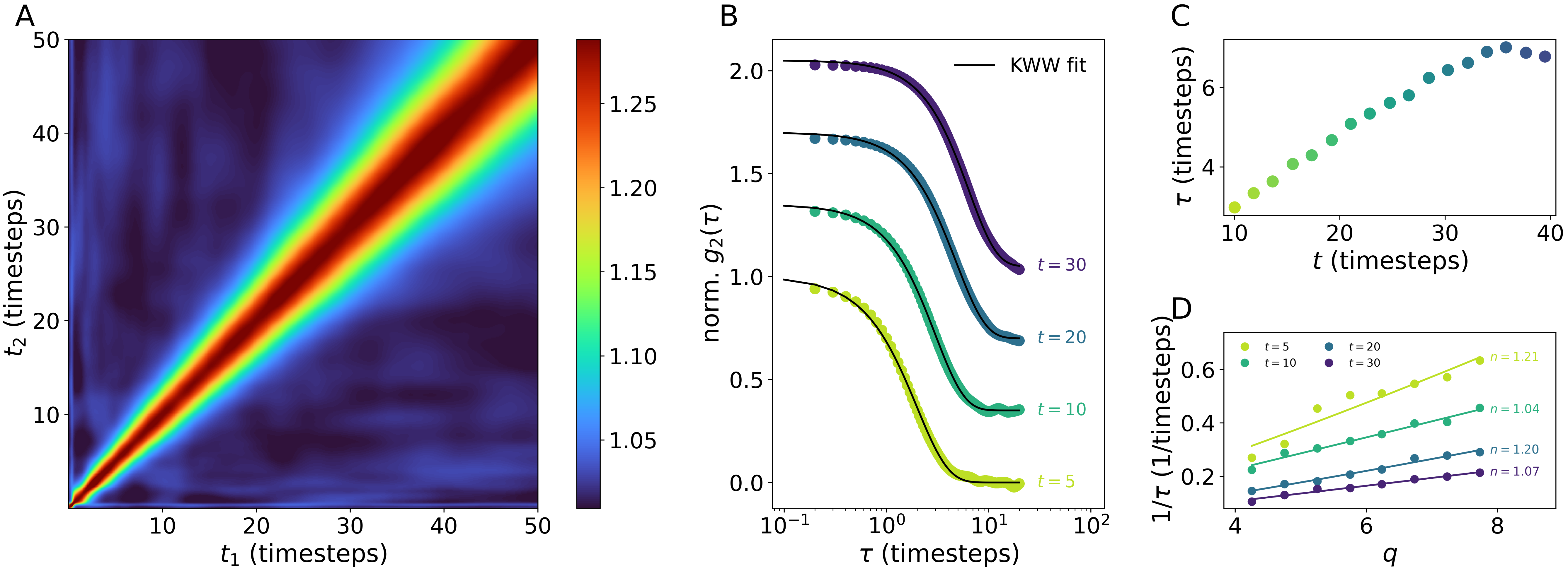}
    \caption{Dynamical evolution obtained from the Cahn-Hilliard (CH) simulations. (A) The two-time correlation (TTC) function at $q=5.75$ (reciprocal simulation units).  
    (B) Intensity autocorrelation functions $g_2(q,t)$ at different waiting times $t$ at $q=5.75$, extracted from the TTC in panel A by averaging horizontally from the main diagonal. The simulated data (data points) are fitted to a KWW function (continuous lines), following the same analysis used for the experimental XPCS data.  The average fitted KWW exponent is $\alpha\approx1.6$. (C) Relaxation times $\tau$ obtained from the KWW fits from panel B. (D) The $q$-dependence of the relaxation rate $1/\tau$. Solid lines depict $1/\tau \propto q^n$ fits with the fitted exponent $n$ indicated in the figure.}\label{fig:CH_ttc}
\end{figure*}

To interpret the observed structural and dynamical evolution, we performed phase-field simulations within the Cahn-Hilliard (CH) framework as a minimal model that contrasts the two canonical pathways of phase separation, spinodal decomposition (SD) and nucleation-and-growth (NG), as described in Sec.~\ref{ss:ch}. We computed the scattering intensity $I(q,t)$ from the simulated order-parameter field, which reflects the evolving spatial correlations of the density field $\phi(\mathbf{r},t)$, i.e.\ the build-up and coarsening of density heterogeneities during SD or NG. The simulated scattering is expressed in reciprocal simulation units and is compared with experiment only on a qualitative basis, since the phase-field model is not calibrated to absolute length scales.\@ Figure~\ref{fig:CH_summary} summarizes the resulting $I(q,t)$ evolution for representative SD and NG simulations, together with simulation snapshots (insets).

To connect more closely to the experimental USAXS analysis (Fig.~\ref{fig:WAXS_saxs}C), we further evaluated the mean intensity $\langle I\rangle_q$ in the $q$-range corresponding to the Porod region (shaded regions in Fig.~\ref{fig:CH_summary}A,B) as a function of simulation time (Fig.~\ref{fig:CH_summary}C). For the SD process, the mean intensity is found to reproduce the experimentally observed mean USAXS intensity trend, rapidly increasing up to a maximum, and then gradually decreasing---whereas the NG pathway yields only a gradual asymptotic increase to a constant value.

Finally, we examine the time evolution of the simulated dynamics, in the SD regime, by computing a two-time correlation function of the spectral intensity (Fig.~\ref{fig:CH_ttc}), analogously to the experimental two-time correlation analysis used for XPCS (Eq.~\ref{c2} and Fig.~\ref{fig:xpcs}). The dynamics slow down progressively with time in a manner that mirrors the experimentally observed gradual increase of the relaxation time. The simulated KWW exponents also exhibit compressed-exponential dynamics ($\alpha\approx1.6$), which are in qualitative agreement with the experiment, although at a different absolute value. A detailed view of the time dependence is shown in the supplementary material, Sec.~S1.1. Moreover, the simulated relaxation rate follows $1/\tau\propto q^n$ with $n=1.21$ at $t=5$ timesteps and $n=1.07$ at $t=30$ timesteps (Fig.~\ref{fig:CH_ttc}D), i.e.\ $n\approx1$ throughout, in qualitative agreement with the exponents obtained from the experimental XPCS data (Fig.~\ref{fig:xpcs}D). We note that the CH model contains neither elasticity nor an internal stress field, so an exponent close to unity arises here purely from the coarsening of the density field. The observation of $n\approx1$ therefore does not by itself call for an interpretation in terms of internal-stress release or dynamical arrest. 

\section{Discussion}\label{s:discussion}

In summary, we measured the structural dynamics of a deeply supercooled water-glycerol solution at $T=172$~K, by a combination of WAXS and USAXS/XPCS.\@ This approach allows us to temporally resolve the liquid structure and dynamics upon cooling deeply in the supercooled region, and to identify the ice crystallite formation based on the emergence of ice Bragg peaks in WAXS.\@ We observe simultaneous, discontinuous changes in the intermolecular liquid structure (WAXS) and the mesoscale structure (from USAXS), as well as in the nanoscale structural dynamics (from XPCS). The latter manifest as a discontinuity in the TTC contrast due to the emergence of nanoscale density fluctuations, indicating a rapid development of mesoscopic heterogeneity in the liquid state. The step-like increase of the Porod exponent to $\approx4$, together with the compressed-exponential dynamics and the hyper-diffusive relaxation rate $1/\tau\propto q$, is consistent with a LLPS scenario accompanied by sharp interface formation. These features are also reproduced by the spinodal-decomposition (SD) branch of our Cahn-Hilliard simulations.
 
A central observation is that ice Bragg peaks appear in WAXS only in stage III, after the structural and dynamical transformation is essentially complete (Figs.~\ref{fig:WAXS_saxs} and~\ref{fig:xpcs}). This timing indicates that the transformation is not driven by crystallization but instead precedes it, implying that a different process is responsible. The data therefore are consistent with an LLPS involving the formation of mesoscopic liquid domains, with ice subsequently nucleating at a later stage.

One possible mechanism underlying such phase separation is an underlying LLT of the kind hypothesized for water and aqueous organic solutions~\cite{murata_liquid_2012, murata_general_2013, woutersen_liquidliquid_2018, tanaka_liquid_2020}. If the deeply supercooled solution can access two liquid states of different local structure, an LLT could provide the thermodynamic driving force for phase separation. In that case, compositional demixing could arise as an aftereffect of the transition between the two liquids. We emphasize, however, that the present data do not allow us to unambiguously determine whether the phase separation is isocompositional~\cite{biddle_behavior_2014} or is accompanied by a net redistribution of glycerol and water. The present WAXS measurements are primarily sensitive to the average liquid structure and do not directly determine the composition of the transient domains. Discriminating between these cases would likely require composition-sensitive probes beyond the structural and dynamical measurements reported here.

An alternative origin is early freeze-concentration, whereby nanocrystallites forming in the bulk expel glycerol from the hydrogen-bond network and locally concentrate the remaining liquid~\cite{filianina_nanocrystallites_2023, bachler_glass_2016}. 
However, no ice Bragg peaks are detected prior to stage III, whereas freeze-concentration would require crystallites to already be present during the structural and dynamical changes of stage II. Moreover, the rapid, simultaneous and discontinuous onset across WAXS, USAXS and XPCS  is difficult to reconcile with the gradual concentration changes expected from freeze-concentration. While we cannot fully exclude a small population of crystallites below our detection limit (see Sec.~\ref{s:results}), freeze-concentration instead provides a natural explanation for the slower, continued shift of the liquid WAXS peak \emph{after} stage III (see Fig.~\ref{fig:setup_protocol}D, inset), once crystallization is established and glycerol is progressively rejected from the growing ice.

Extending the approach presented here to different solute concentrations and quench temperatures, particularly towards lower glycerol concentrations, would help discriminate between LLPS driven primarily by water-glycerol demixing, and LLPS driven by an underlying liquid-liquid transition, and would quantify how the resulting heterogeneity influences subsequent ice nucleation. It would further be interesting to explore the samples under elevated pressure and probe the possible existence of a liquid-liquid critical point in these solutions~\cite{suzuki_experimentally_2014, suzuki_experimental_2018}.

\section*{Supplementary Material}
See the supplementary material for the time dependence of the $q$-averaged KWW exponent for the experiment and simulations, the details of the power-law fits to the $q$-dependence of the relaxation rate, a repeat experiment on a fresh sample quenched to $T=182$~K (WAXS, USAXS, and XPCS), and the detector mask used for the XPCS correlation analysis together with the resulting speckle contrast.

\section*{Acknowledgements}
Parts of this research were carried out at P10 beamline (proposal I-20220280) at the light source PETRA III at DESY, a member of the Helmholtz Association (HGF). We also acknowledge the European Synchrotron Radiation Facility (ESRF) for provision of synchrotron radiation facilities at ID10 beamline (proposal SC5359). F.P. acknowledges financial support by the Swedish National Research Council (Vetenskapsrådet) under Grant Nos. 2019-05542 and 2023-05339, and within the Röntgen-Ångström Cluster Grant No. 2019-06075, and the kind financial support from the Knut och Alice Wallenberg Foundation (WAF, Grant No. 2023.0052). This research is supported by the Center of Molecular Water Science (CMWS) of DESY in an Early Science Project, the MaxWater initiative of the Max-Planck-Gesellschaft, Carl Tryggers (Project No. CTS21:1589) and the Wenner-Gren Foundations (Project No. UPD2021-0144). F.L. is supported by the Deutsche Forschungsgemeinschaft (DFG, German Research Foundation) as part of the Excellence Strategy of the Federal Government and the federal states – EXC 3120/1 BlueMat: Water-Driven Materials – project number 533771286. Finally, F.P. and I.A. acknowledge funding from the European Union's Horizon Europe research and innovation programme under the Marie Skłodowska-Curie grant agreement No. 101081419. 

\section*{AUTHOR DECLARATIONS}
\subsection*{Conflict of Interest}
The authors have no conflicts to disclose.

\section*{Data Availability}
Raw data were generated at the PETRA III (DESY) and ESRF large-scale facilities. The data that support the findings of this study are available from the corresponding author upon reasonable request.

\bibliography{references_trimmed}

\clearpage
\onecolumngrid
\begin{center}
{\large\textbf{Supplementary Material}}\\[6pt]
{\large\textbf{Liquid-liquid phase separation precedes crystallization in supercooled water-glycerol solutions}}
\end{center}
\vspace{1.5em}

\setcounter{section}{0}
\setcounter{subsection}{0}
\setcounter{figure}{0}
\setcounter{table}{0}
\setcounter{equation}{0}
\renewcommand{\thesection}{S\arabic{section}}
\renewcommand{\thesubsection}{S\arabic{section}.\arabic{subsection}}
\renewcommand{\theequation}{S.\arabic{equation}}
\renewcommand{\thefigure}{S\arabic{figure}}
\renewcommand{\thetable}{S\Roman{table}}
\renewcommand{\theHsection}{S\arabic{section}}
\renewcommand{\theHfigure}{S\arabic{figure}}
\renewcommand{\theHtable}{S\Roman{table}}
\renewcommand{\theHequation}{S.\arabic{equation}}
\section{Fits to the XPCS correlation functions}

\subsection{KWW exponent for experiment and simulations}

\begin{figure}[htb]
\centering
\includegraphics[width=\textwidth]{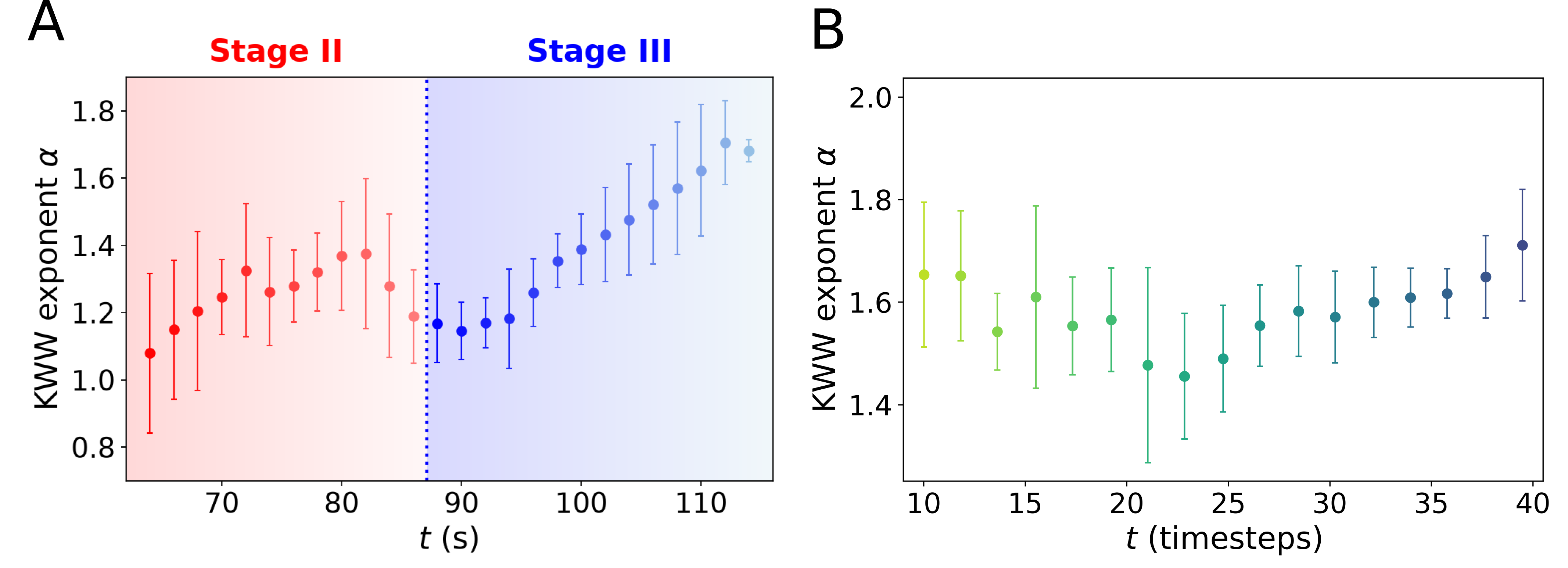}
\caption{\label{fig:kww_exp}
$q$-averaged KWW exponent $\alpha$ versus waiting time $t$. (A) Experiment, separated into stage~II (red) and stage~III (blue). (B) Cahn-Hilliard simulations (simulation time in time steps). Error bars denote the standard error of the $q$-average.
}
\end{figure}

The shape of the relaxation is quantified by the KWW exponent $\alpha$ obtained from the $g_2$ fits (see the main text). Since $\alpha$ is essentially independent of the momentum transfer $q$, we report its $q$-averaged value as a function of the waiting time $t$; error bars denote the standard error of the $q$-average. Figure~\ref{fig:kww_exp}A shows the experimental result and Fig.~\ref{fig:kww_exp}B that of the Cahn-Hilliard (CH) simulations.

In the experiment (Fig.~\ref{fig:kww_exp}A), $\alpha$ remains roughly constant at $\alpha\approx1.1-1.2$ throughout stage~II and rises towards $\alpha\approx1.5$ during stage~III. In the CH simulations (Fig.~\ref{fig:kww_exp}B), $\alpha$ is similarly roughly constant at early times before increasing at later times. The two show qualitatively similar behavior: a nearly constant exponent during the transformation followed by an increase at later times, although their absolute values differ. In both cases $\alpha$ is predominantly greater than unity, corresponding to compressed-exponential relaxation, consistent with the ballistic-like dynamics ($1/\tau\propto q$) discussed in the main text.

\subsection{\texorpdfstring{Power-law fits to the $q$-dependence of the relaxation rate}{Power-law fits to the q-dependence of the relaxation rate}}\label{sec:qdep}

The $q$-dependence of the relaxation rate $\Gamma=1/\tau$ shown in Fig.~3D of the main text is described by a free power law $\Gamma(q)=Aq^{n}$, with both the amplitude $A$ and the exponent $n$ treated as free parameters. The fits are unweighted and performed on the untransformed $\Gamma$ values, with the uncertainty on $n$ estimated from the residuals of each fit. For each stage we quote the mean of the per-waiting-time exponents, with the spread between waiting times as the uncertainty; three waiting times fall in stage~II ($t=64$, $72$ and $79$~s) and two in stage~III ($t=104$ and $109$~s).

The innermost ROI ($q=0.011$~nm$^{-1}$) is excluded from the exponents quoted in the main text, as it contains an order of magnitude fewer detector pixels than the others and carries the least stable speckle contrast of the six (Sec.~S3). Fitting over all six ROIs gives $n=1.10\pm0.16$ in stage~II and $n=0.82\pm0.06$ in stage~III; excluding the innermost ROI gives $n=1.08\pm0.13$ and $n=1.01\pm0.08$, the values quoted in the main text. In either case the exponents of the two stages coincide within their uncertainties.

\clearpage
\section{\texorpdfstring{Quenching to $T=182$~K}{Quenching to T=182 K}}\label{sec:S2}
\begin{figure}[htb]
\centering
 \includegraphics[width=0.5\textwidth]{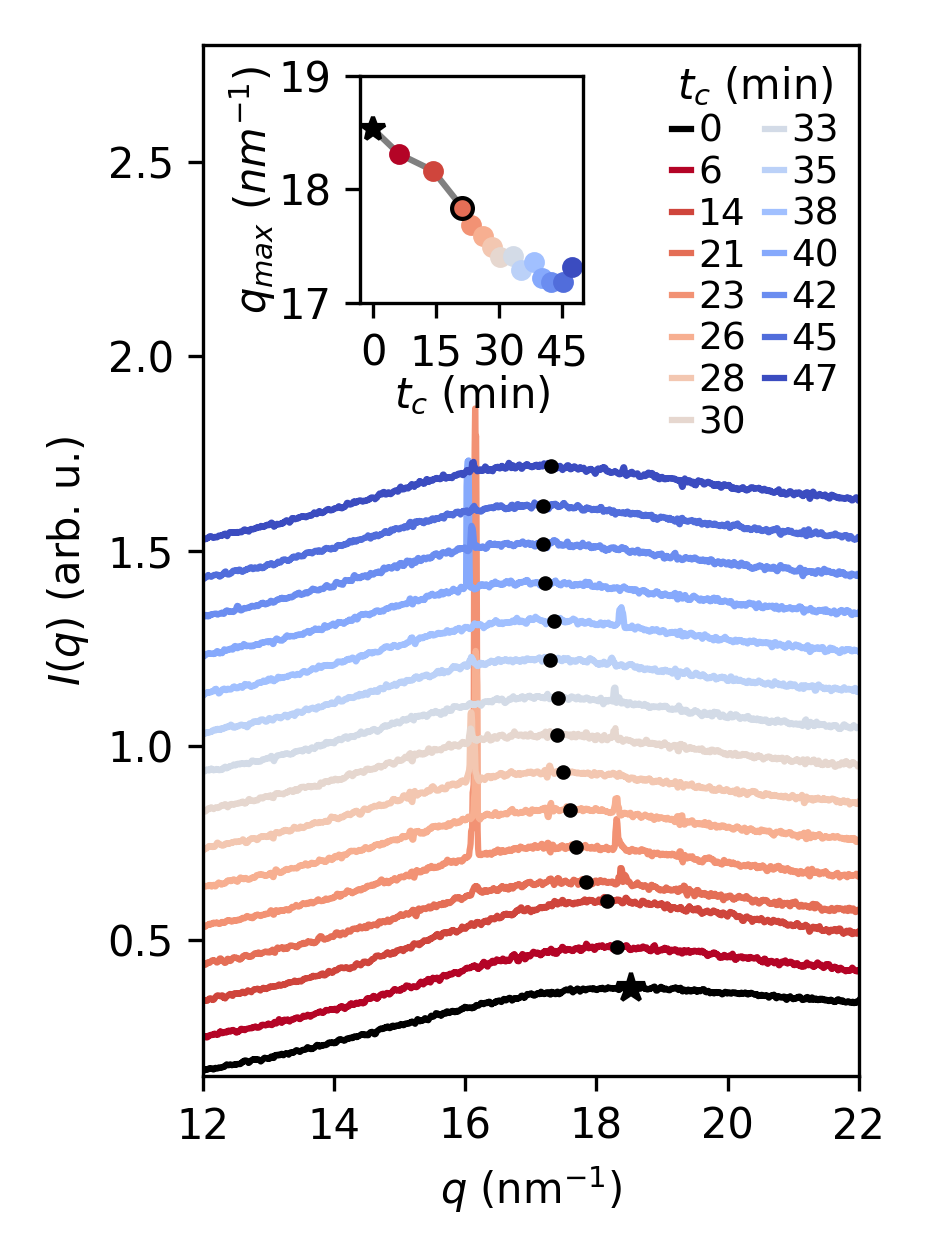}
 \caption{Wide-angle X-ray scattering (WAXS) of 16.5~mol\% water-glycerol solution upon quenching to $T_{quench}=182\pm6$~K. Here, each curve represents the WAXS pattern measured at a different sample spot, with cooling time counted from the start of the quench at a cooling rate 20~K/min. A measurement at room temperature $T=295$~K prior to the quench is shown in black. The inset shows the liquid peak position $q_{\max}$ in each spot versus the cooling time, where the black star and black circle indicate the peak position at room temperature and the measurement at 21~min of cooling. A relative vertical shift of the WAXS patterns has been introduced for clarity.
 }\label{fig:waxs_spots_170}
\end{figure}

\begin{figure}[htb]
\centering
 \includegraphics[width=\textwidth]{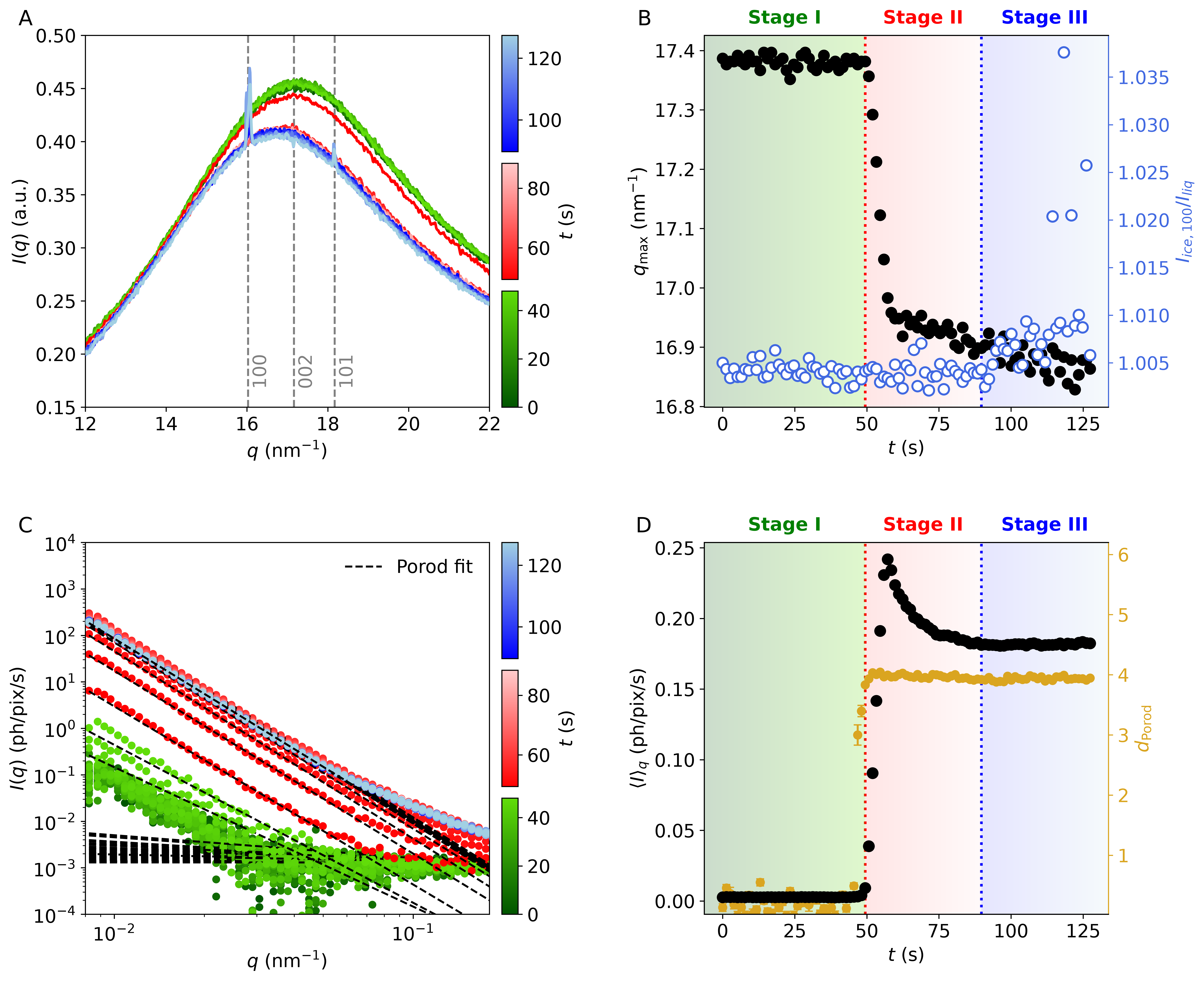}
 \caption{Simultaneous WAXS (A, B) and USAXS (C, D) measurements during the structural transformation after quenching to $T_{quench}=182\pm6$~K, at 21~min of cooling. Three distinct stages are identified: a stable supercooled liquid (stage I, green), a rapid structural transformation (stage II, red), and the onset of ice crystallization (stage III, blue). (A) WAXS patterns $I(q)$ versus waiting time $t$ (colorbar); vertical dashed lines mark the (100), (002) and (101) Bragg peaks of hexagonal ice. (B) Evolution of the liquid peak position $q_{\max}$ (black circles, left axis) and the intensity ratio $I_{ice,100}/I_{liq}$ of the ice (100) Bragg peak at $q=16.13$~nm$^{-1}$ to the liquid peak (open blue circles, right axis) versus $t$. The dotted lines and shaded regions indicate the boundaries between the different stages. (C) USAXS patterns $I(q)$ versus $t$. Black dashed lines show power-law (Porod) fits to Eq.~(1) of the main text. (D) Mean scattering intensity $\langle I\rangle_q$ averaged over $0.02\leq q\leq0.3$~nm$^{-1}$ (black, left axis) and the fitted Porod exponent $d_{Porod}$ from the fits in panel~C (gold, right axis) versus $t$.
 }\label{fig:waxs_saxs_170}
\end{figure}

\begin{figure}[htb]
\centering
 \includegraphics[width=\textwidth]{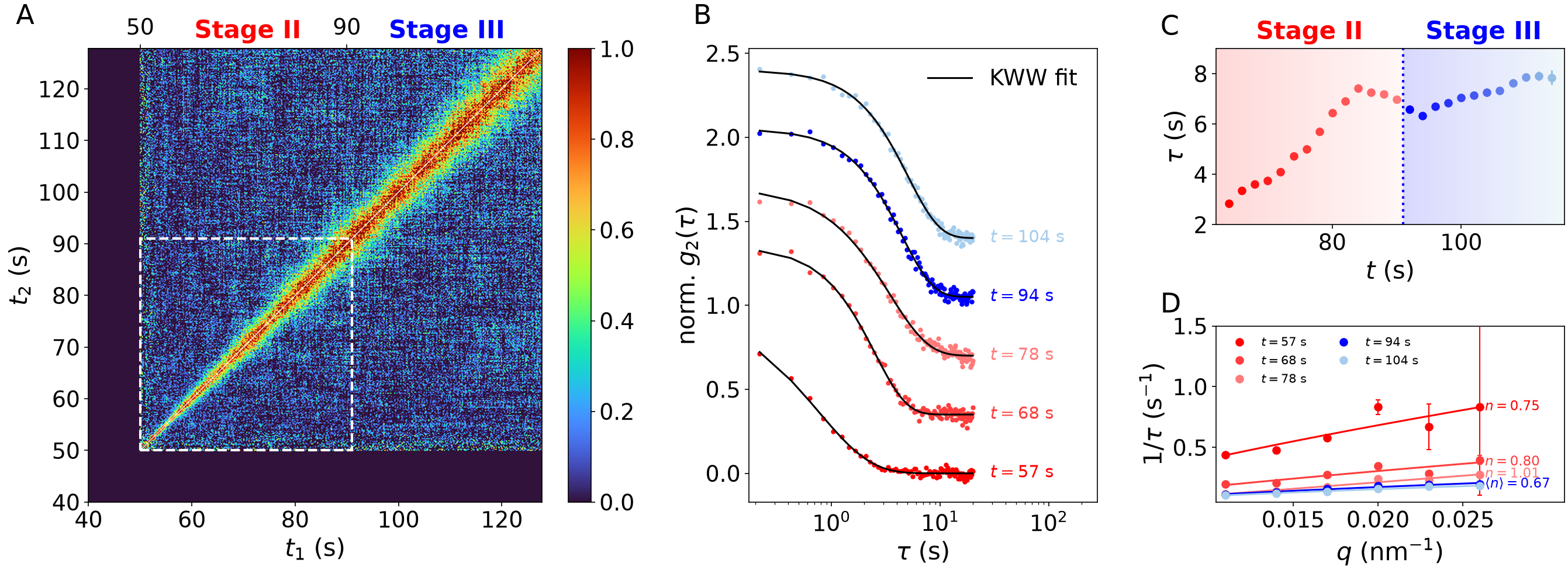}
 \caption{X-ray photon correlation spectroscopy (XPCS) after quenching to $T_{quench}=182\pm6$~K. (A) The two-time intensity correlation function (TTC) measured at 21~min of cooling and $q=0.016$~nm$^{-1}$. The stage boundaries obtained from the simultaneous WAXS and USAXS measurements (Fig.~\ref{fig:waxs_saxs_170}) are marked on the top axis, and the white dashed square outlines stage~II. (B) The intensity correlation functions $g_2$ for different waiting times $t$ at $q=0.016$~nm$^{-1}$, extracted from the TTC in panel A by binning the correlation signals from the main diagonal ($t_1=t_2$) along the $t_1$-axis. The experimental data (dots) are fitted to Kohlrausch-Williams-Watts (KWW) functions (lines) and normalized by the offset $c$ and contrast $\beta$. (C) The average relaxation time $\tau$ versus the waiting time $t$, averaged over $0.013\leq q\leq 0.019$~nm$^{-1}$. (D) The $q$-dependence of the inverse relaxation time $1/\tau$ obtained from the fits to the $g_2$ in panel B. Solid lines depict $1/\tau\propto q^n$ fits over all six ROIs, with the fitted exponent $n$ indicated in the figure; for stage III the average $\langle n\rangle$ over the two waiting times is given, as the data points overlap (see Sec.~S1.2). Dynamics in B-D extracted before and after the appearance of ice peaks in WAXS ($t\approx90$~s) are indicated in red and blue shades respectively.
 }\label{fig:xpcs_170}
\end{figure}

We additionally performed measurements on a different sample of 16.5~mol\% water-glycerol solution upon quenching to $T_{quench}=182\pm6$~K. The results are summarized in Figs.~\ref{fig:waxs_spots_170}-\ref{fig:xpcs_170}. As with the lower-quench-temperature measurements, WAXS reveals a discontinuous shift in the liquid peak position (at 21~min of cooling) that precedes detectable ice crystallization ((see Figs.~\ref{fig:waxs_spots_170} and~\ref{fig:waxs_saxs_170}A, B). This shift occurs simultaneously with a discontinuous enhancement of the USAXS intensity (Fig.~\ref{fig:waxs_saxs_170}C, D) and a sudden increase in contrast along the diagonal of the two-time correlation (TTC) function at $t\approx52$~s (Fig.~\ref{fig:xpcs_170}A), consistent with the LLPS scenario discussed in the main text. We also find that the evolution of the solution dynamics is largely reproduced, with a gradual slowdown (Fig.~\ref{fig:xpcs_170}B-D, red shades) followed by a plateau after the onset of ice crystallization (blue shades, Fig.~\ref{fig:xpcs_170}B-D). The relaxation is compressed-exponential throughout ($\alpha\approx1.3-1.5$), and power-law fits to the $q$-dependence of the relaxation rate, performed over all six ROIs as described in Sec.~S1.2, give $n=0.85\pm0.14$ in stage~II and $n=0.67\pm0.02$ in stage~III.

\clearpage
\section{Detector mask and speckle contrast of the XPCS analysis}\label{sec:mask}

\begin{figure}[htb]
\centering
 \includegraphics[width=\textwidth]{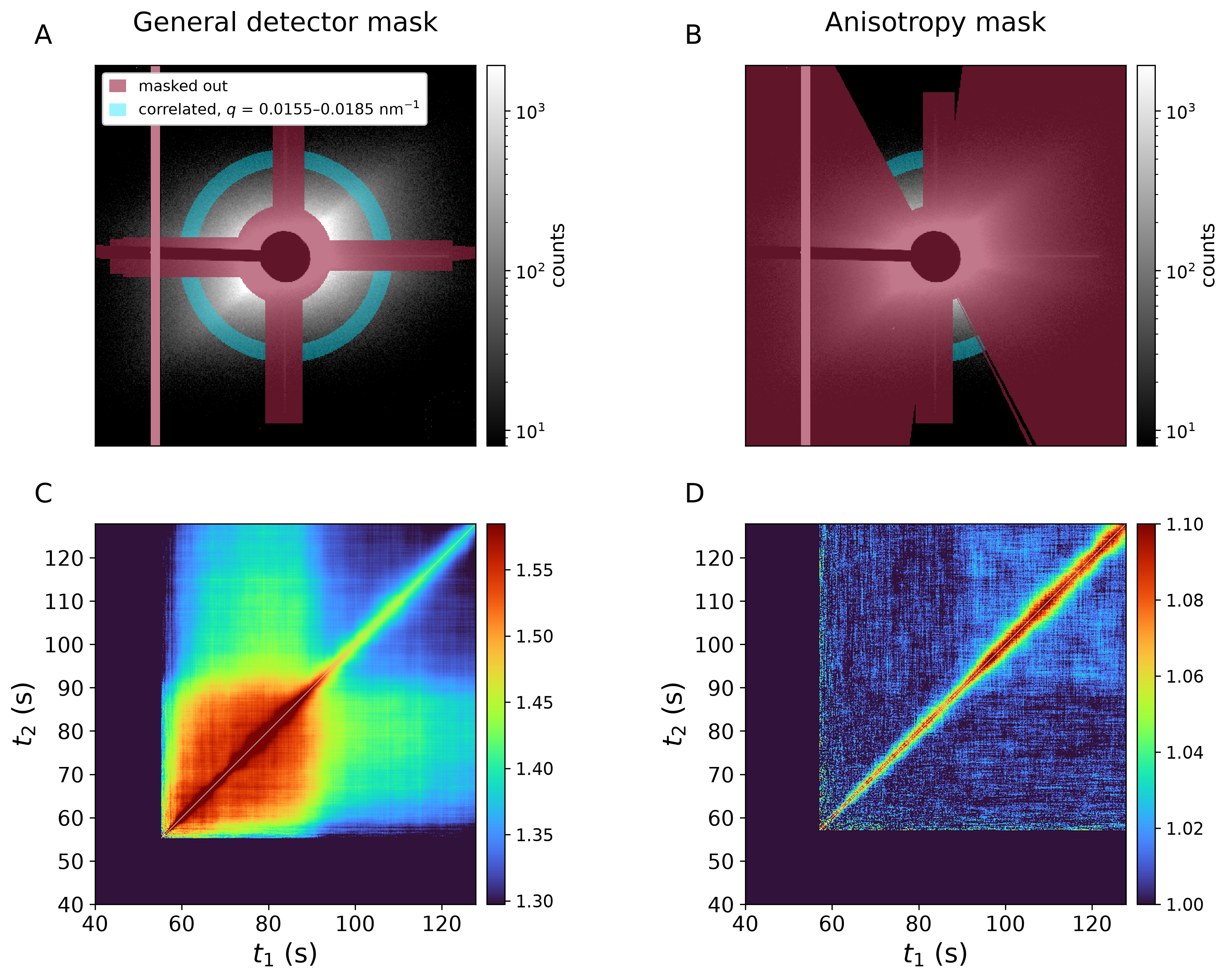}
 \caption{Effect of the detector mask on the XPCS correlation, for the measurement discussed in the main text ($T_{quench}=172\pm6$~K, 25~min of cooling). Left column (A, C): general detector mask. Right column (B, D): anisotropy mask of this sample. Only the mask differs between the two columns. (A, B) The detector image seen by the correlation (80 summed raw frames, logarithmic gray scale), with the pixels discarded by the mask shown in dark red and the pixels of the displayed ROI ($q=0.0155-0.0185$~nm$^{-1}$) that entered the correlation in cyan. (C, D) The two-time correlation function of that ROI, in raw $c_2$ values. Note the different color scales, $1.30-1.59$ in C and $1.00-1.10$ in D. The uniform dark region at low $t_1$ and $t_2$ is where the correlation is undefined.
 }\label{fig:mask_172}
\end{figure}

\begin{figure}[htb]
\centering
 \includegraphics[width=\textwidth]{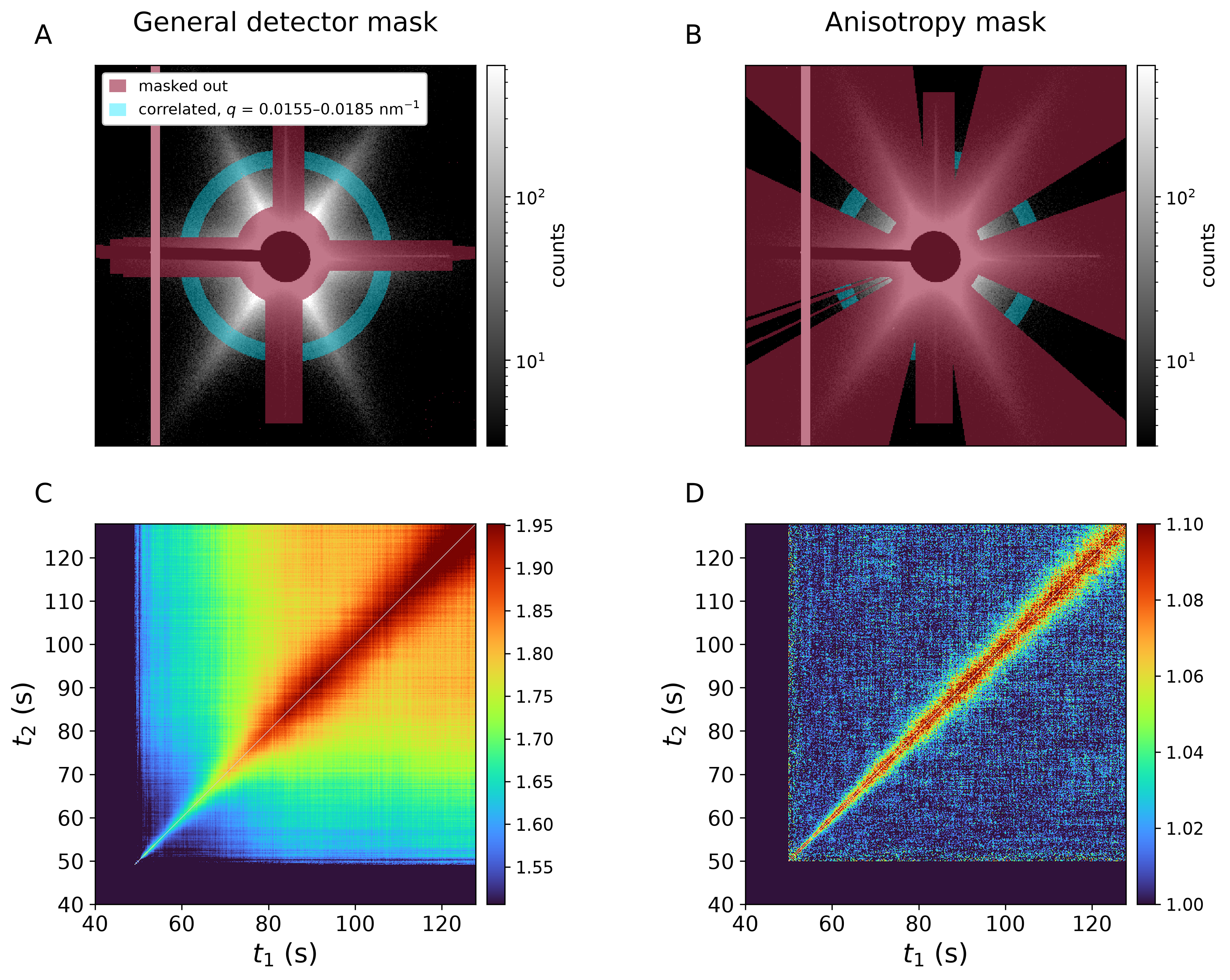}
 \caption{Effect of the detector mask on the XPCS correlation, for the repeat measurement quenched to $T_{quench}=182\pm6$~K (Sec.~S2, 21~min of cooling); layout as in Fig.~\ref{fig:mask_172}. Left column (A, C): general detector mask. Right column (B, D): anisotropy mask of this sample. (C, D) The two-time correlation function of the displayed ROI ($q=0.0155-0.0185$~nm$^{-1}$), in raw $g_2$ values. Note the different color scales, $1.52-1.98$ in C and $1.00-1.10$ in D. The uniform dark region at low $t_1$ and $t_2$ is where the correlation is undefined.
 }\label{fig:mask_182}
\end{figure}

The pixel average of the two-time correlation function [Eq.~(2) of the main text] runs over the detector pixels of an annulus of constant $q$, but not every pixel of such an annulus is usable. Figures~\ref{fig:mask_172} and~\ref{fig:mask_182} compare, for the measurement discussed in the main text and for the repeat quenched to $T_{quench}=182\pm6$~K of Sec.~S2, the correlation obtained with the general detector mask, which keeps every pixel that is not a module gap, a beamstop shadow or a bad pixel, with the those obtained using the sample-specific anisotropy masks employed in our analysis.

Raw frames ($0.0128~$s) were summed in bins of 16 into frames of $0.2048~$s. From 10,000 frames acquired, the first 2000 were excluded (low signal; pre-equilibration phase), yielding 500 usable frames covering waiting times $t = 25.6$–$127.8$~s. Six $q$-ROIs are defined, with centers from $q=0.011$ to $0.026$~nm$^{-1}$ in steps of $0.003$~nm$^{-1}$ and a half-width of $0.0015$~nm$^{-1}$. The ROI displayed in Figs.~\ref{fig:mask_172} and~\ref{fig:mask_182} is the third of these, $q=0.0155-0.0185$~nm$^{-1}$, which contains the $q=0.016$~nm$^{-1}$ correlation functions shown in the main text and in Fig.~\ref{fig:xpcs_170}. Correlating a pair of frames requires at least 64 illuminated pixels in both; frames that are too weakly illuminated are discarded as a whole, leaving the uniform dark region at low $t_1$ and $t_2$ in Figs.~\ref{fig:mask_172}C, D and~\ref{fig:mask_182}C, D.

As seen in Figs.~\ref{fig:mask_172}A and~\ref{fig:mask_182}A, the small-angle scattering of these samples is strongly anisotropic, with the small-angle ring crossed by pronounced streaks of largely static intensity, whose origin we do not identify here. A $q$-ROI defined using the general detector mask forms a complete annulus and therefore intersects these streaks. Because Eq.~(2) of the main text normalizes by the pixel-averaged intensity at each time, a time-independent contribution to $I(q,t)$ adds a lag-independent offset to $c_2$ instead of decaying together with the dynamics of the sample. Averaged over the stage~III waiting times of the displayed ROI, the correlation obtained with the anisotropy mask decays to the baseline of unity assumed in Eq.~(4) of the main text, whereas with the general detector mask 73\% (172~K) and 84\% (182~K) of the excess over unity at the shortest lag does not decay within the measured time window. Normalizing the $g_2$ curves by such an apparent contrast would bias both $\tau$ and $\alpha$. The streak pixels are not perfectly static, as their intensity follows the overall trend of the pixel-averaged intensity, and they are masked in full so that this contribution stays separate from the dynamics of interest.

The anisotropy mask is drawn separately for each sample by removing the azimuthal sectors that carry the streaks, and in each case keeps a strict subset of the pixels kept by the general detector mask. At 172~K it retains 17.5\% of these pixels, leaving two opposed wedges of the ROI annulus, and at 182~K it retains 27.1\%, leaving six narrower wedges (Figs.~\ref{fig:mask_172}B and~\ref{fig:mask_182}B). Fitting Eq.~(4) of the main text per ROI at every waiting time then yields $\beta=0.0824\pm0.0044$ (172~K) and $\beta=0.0887\pm0.0015$ (182~K), where the uncertainty is the spread across the six ROIs, and the stage-II and stage-III means agree within that spread for every ROI. The contrast is least stable in the innermost ROI at 172~K, where it drifts downwards over the scan, which is why the power-law fits used to obtain the exponents quoted in the main text exclude the innermost ROI (Sec.~S1.2). 
The two measurements yield contrast values that agree within 7\% despite the different masks, as expected for a contrast determined primarily by the beam coherence and the ratio of speckle size to pixel size rather than by the sample. In contrast, using the general detector mask yields values differing by approximately a factor of two, reflecting the different contributions from static scattering in the two measurements.

\end{document}